\documentclass[journal]{IEEEtran}

\ifCLASSINFOpdf
\else
   \usepackage[dvips]{graphicx}
\fi
\usepackage{url}
\usepackage{multirow}
\usepackage{graphicx}
\usepackage{amsmath}
\usepackage{amssymb}
\usepackage{dsfont}
\usepackage{graphicx}

\begin{document}

\title{Graph-based Interaction Augmentation Network for Robust Multimodal Sentiment Analysis}

\author{Zhangfeng Hu, Mengxin Shi, \thanks{Zhangfeng Hu is with School of Biological Science and Medical Engineering, Southeast University, Nanjing 210096, China (email: zhangfeng hu@seu.edu.cn). Mengxin Shi is with School of Information Science and Engineering, Southeast University, Nanjing 210096, China (email: shimengxin@seu.edu.cn).}}

\markboth{Journal of \LaTeX\ Class Files, Vol. 14, No. 8, August 2015}
{Shell \MakeLowercase{\textit{et al.}}: Bare Demo of IEEEtran.cls for IEEE Journals}
\maketitle

\begin{abstract}
The inevitable modality imperfection in real-world scenarios poses significant challenges for Multimodal Sentiment Analysis (MSA). While existing methods tailor reconstruction or joint representation learning strategies to restore missing semantics, they often overlook complex dependencies within and across modalities. Consequently, they fail to fully leverage available modalities to capture complementary semantics. To this end, this paper proposes a novel graph-based framework to exploit both intra- and inter-modality interactions, enabling imperfect samples to derive missing semantics from complementary parts for robust MSA. Specifically, we first devise a learnable hypergraph to model intra-modality temporal dependencies to exploit contextual information within each modality. Then, a directed graph is employed to explore inter-modality correlations based on attention mechanism, capturing complementary information across different modalities. Finally, the knowledge from perfect samples is integrated to supervise our interaction processes, guiding the model toward learning reliable and robust joint representations. Extensive experiments on MOSI and MOSEI datasets demonstrate the effectiveness of our method \footnote{https://github.com/flotaas/GIAN}. 
\end{abstract}

\IEEEpeerreviewmaketitle

\section{Introduction}
\IEEEPARstart{P}{revious} research \cite{tsai2019multimodal,hazarika2020misa,li2023decoupled} on Multimodal Sentiment Analysis (MSA) has demonstrated promising performance under the assumption that all modalities are available by exploiting the common and complementary cues among different modalities, including visual, audio, and language, to understand the speakers' affective state. However, in real-world scenarios, it is challenging to obtain perfect modality data due to factors such as privacy concerns, device malfunctions, and background noise. Fig.1 depicts three types of modality imperfection, which inevitably degrade the effectiveness of well-trained MSA models in real-world applications.     

\begin{figure}[h]
    \centering
    \includegraphics[width=0.9\linewidth]{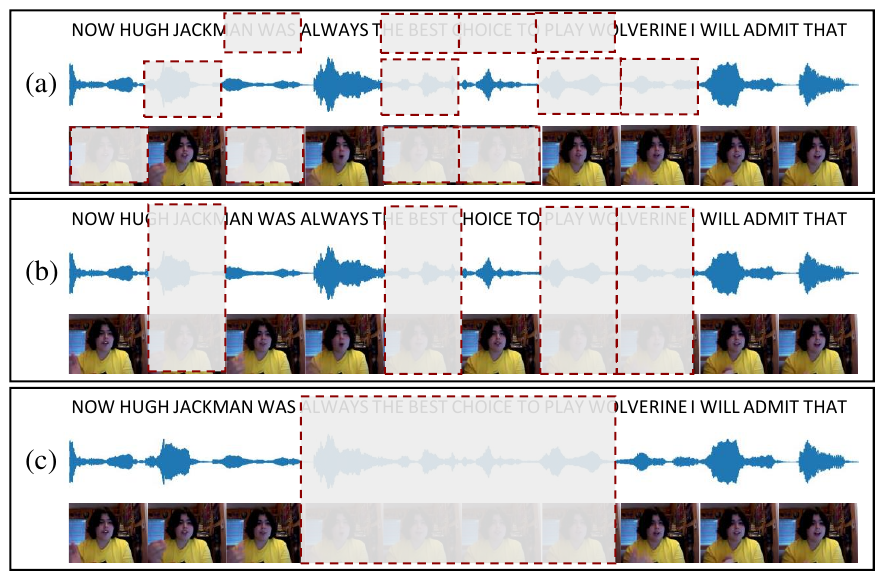}
    \caption{Illustration of three types of modality imperfection cases: (a) \textbf{Random Missing}, where features are randomly missing within each modality respectively; (b) \textbf{Temporal Missing}, where features are simultaneously missing across all modalities at random positions; (c) \textbf{Structural temporal Missing}, where features are simultaneously, consecutively missing across modalities.} 
    \label{demo}
\end{figure}

In recent years, extensive research efforts \cite{zhao2021missing,zeng2022tag,wang2023distribution,xie2023exploring,wang2024incomplete} have been proposed toward developing robust MSA frameworks to address modality imperfection. There are two principle strategies: 1) Reconstruction for specific modality representations, which aims to explicitly generate the features of missing modality based on available modalities. \cite{lin2021completer} restored missing modality from the available one by maximizing mutual information between the two. \cite{lian2023gcnet} introduced relation graph to reconstruct missing modality based on temporal and speaker dependencies across modalities. \cite{wang2024incomplete} exploited the score-based diffusion model to restore the distribution of missing modalities. Despite offering high interpretability, ensuring the quality of the generated missing modalities remains challenging, and inaccurate reconstruction may propagate to downstream tasks, potentially degrading MSA performance. 2) Joint representation learning focuses on directly learning a high-level joint representation across incomplete modalities without seperately generating each missing modality. Generally, multimodal knowledge is distilled to recover missing joint semantics. \cite{li2023towards} devised a self-distillation framework to transfer beneficial knowledge to incomplete samples for missing semantics recovery. \cite{yuan2023noise} integrated adversarial learning to narrow the knowledge gap between the representations of perfect and imperfect pairs. While effective, these methods only account for simple relationships between modalities, overlooking the complex, higher-order dependencies that span both intra- and inter-modality dimensions. Consequently, they fail to fully leverage available modalities to derive robust semantics.

Thus, in this paper, we propose a novel framework for robust MSA, termed Graph-based Interaction Augmentation Network (GIAN). Our basic idea is that complementary information exists both within and across modalities, and missing semantics can be derived from complementary parts to learn robust representations. Thus, GIAN aims to sufficiently exploit interactions within imperfcet samples to mine complementary semantics. Specifically, by leveraging the effective information propagation of graphs, we propose two graph-based modules, i.e., Learnable temporal Hypergraph Module (LTHM) and Attention-based Multimodal Graph Module (AMGM), to exploit temporal intra-modality interaction and semantic intermodality interaction respectively. Fisrt, by transforming the sequence features of each modality into a learnable hypergraph, LTHM explores contextual dynamics and captures temporal dependencies within modalities. Then, by modeling multimodal semantic dependancies with a directed graph, AMGM explores multimodal interaction and aggregates complementary information among modalities via attention mechanism. Finally, a feature refinement strategy is proposed to supervise our interaction processes, guiding the model towards learning reliable joint representations under modality imperfection. Our contributions are three-fold: 1) We propose a novel graph-based framework for robust MSA, which expoits both intra- and inter-modality interactions within imperfect samples to derive complementary semantics; 2) We model temporal intramodality and semantic inter-modality dependencies with a learnable hypergraph and a directed graph respectively, which ensures complex high-order relations within and across modalities are not neglected; 3) Extensive experiments on two MSA benchmark datasets indicate that the proposed GIAN consistently achieves significant improvements on MSA across various imperfect data scenarios.

\vspace{-0.3cm}
\begin{figure*}[h]
    \centering
    \includegraphics[width=0.7\linewidth]{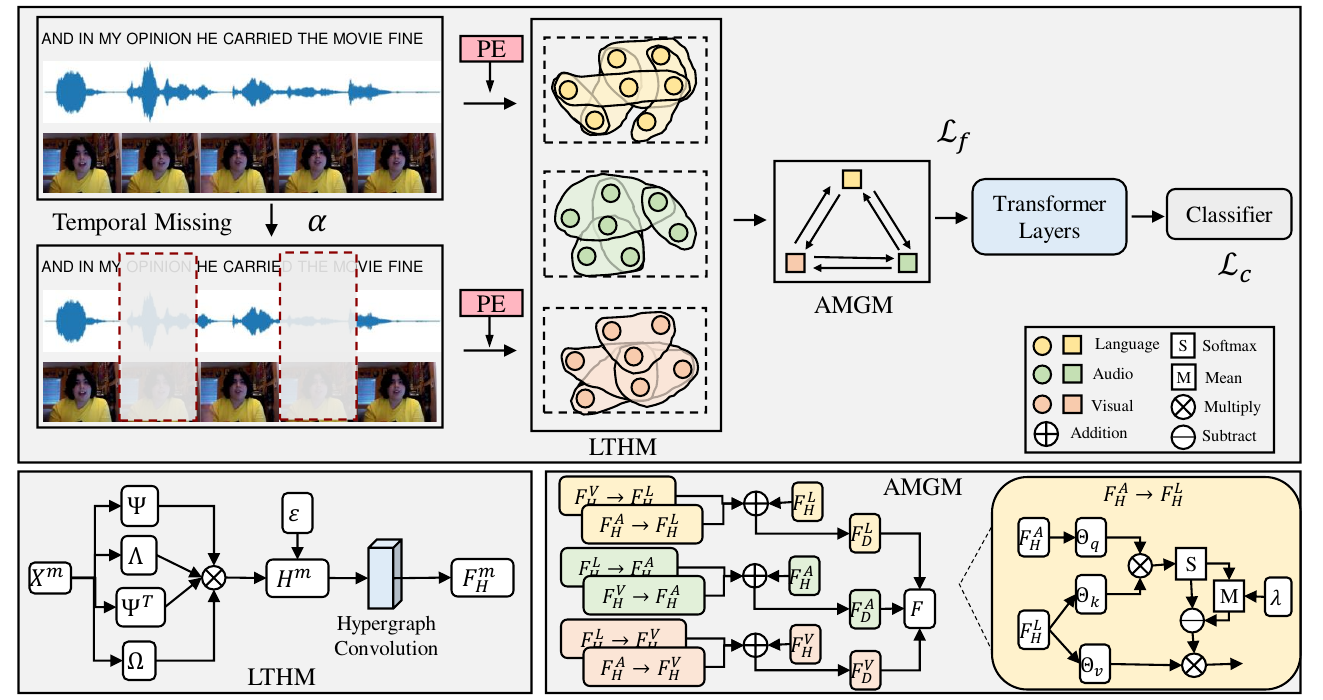}
    \caption{The framework of our GIAN. The multimodal data are first processed through modality-specific LTHM to explore intra-modality temporal dependencies. Then, AMGM is employed to capture inter-modality relations and interactions. Finally, the concatenated multimodal features are fed into Transformer layers for fusion and passed through a classifier for prediction. A feature refinement strategy is introduced to supervise this process for robust representation learning.} 
    \label{framework}
\end{figure*}

\section{Proposed Method}
\subsection{Problem Formulation}
Given a multimodal dataset $\text{D}=\{\text{D}^\text{train}, \text{D}^\text{val}, \text{D}^\text{test}\}$, each sample $S$ consists of three modalities, i.e., vision, audio and language, denoted as $S =\{\boldsymbol{X}^{V}, \boldsymbol{X}^{A}, \boldsymbol{X}^{L}\}$, respectively. $\boldsymbol{X}^{m} \in \mathbb{R}^{T \times d_{m}}, m \in \{V, A, L\}$, $T$ is the sequence length and $d_m$ is the feature dimension. During the training stage, we assume the complete multimodal data is avaliable. To learn robust joint representations, we train the model using imperfect samples to explore both intra- and inter-modality interactions to capture complementary semantics. Specifically, we define an imperfect ratio $\alpha \in (0, 1)$, and randomly set $\alpha$ temporal features of all modalities to zero to obtain temporal missing version of $S$, donated as $S^{\prime} =\{\boldsymbol{X}_{V}^{\prime}, \boldsymbol{X}_{A}^{\prime}, \boldsymbol{X}_{L}^{\prime}\}$. Then, the trained model is evaluated under three types of modality imperfection using disrupted $\text{D}^\text{val}$ and $\text{D}^\text{test}$. 

The framework of our GIAN is depicted in Fig.2, which includes two modules: LTHM and AMGM. Thanks to the effective information propagation of graphs, they can derive complementary semantics via intra- and inter-modality interactions from imperfect samples respectively.

\vspace{-0.2cm}
\subsection{Learnable temporal Hypergraph Module (LTHM)}
In this module, we transform sequence features of each modality into a learnable hypergraph, using hypergraph convolution to capture the long-term temporal dependency within modalities. Compared to conventional graphs, which can only model pairwise relationships, hypergraphs are capable of connecting multiple nodes through hyperedges, allowing them to capture higher-order connections across time, ensuring that complex contextual patterns are not fragmented or neglected. 

\textbf{Positional Encoding.} Before constructing the hypergraph, we first apply a positional encoding \cite{vaswani2017attention} to each sequence feature to avoid lossing temporal information. This is denoted as $\boldsymbol{X}^{m}_t = \boldsymbol{X}^{m}_t + P E_t$, where $\boldsymbol{X}^{m}_t \in \mathbb{R}^{1 \times d_{m}}$ is $t^{th}$ sequence feature of modality $m\in \{V,A,L\}$, $ {PE}_t^{(i)}= \sin \left(\frac{1}{10000^{2 k / d_m}} t\right)$, if $i=2k$, and $ {PE}_t^{(i)}=\cos \left(\frac{1}{10000^{2 k / d_m}} t\right)$, if $i=2 k+1$, $i$ is the dimension. 

\textbf{Hypergraph Construction and Convolution.} For each modality $m$, we consider each element in the sequence as a node with the feature vector $\boldsymbol{X}^{m}_t$ to construct a hypergraph. Let $\mathcal{G}^m=(\mathcal{V}^m, \mathcal{E}^m, \mathbf{W}^m, \mathbf{H}^m)$ denote a hypergraph with node set $\mathcal{V}^m=\{\boldsymbol{X}^{m}_1, \boldsymbol{X}^{m}_2,..., \boldsymbol{X}^{m}_{T_m}\}$, hyperedge set $\mathcal{E}^m=\{e^m_1, e^m_2, ..., e^m_M\}$, and corresponding weight matrix of hyperedges $\mathbf{W}^m \in \mathbb{R}^{M \times M}$, $M$ denotes the number of hyperedges. The adjacency matrix $\mathbf{H}^m \in \mathbb{R}^{T_m \times M}$ depicts the relationship between nodes and hyperedges, which is then used to propagate information among nodes. Previous methods \cite{jiang2019dynamic,huang2024dynamic} employ KNN clustering to construct hypergraph structures by connecting each node to its $K$ nearest neighbors in the feature space, thereby forming a hyperedge that contains $K + 1$ nodes. However, it suffers from two major limitations. First, the fixed number of neighbors ($K$) may not adapt well to the non-uniform distribution of data, potentially missing important connections in dense regions while introducing irrelevant nodes in sparse areas. Second, the KNN-based method primarily focuses on local structures and may overlook global high-order relationships, thereby limiting the expressiveness of the hypergraph in modeling complex structures. Inspired by \cite{wadhwa2021hyperrealistic,qiu2024high}, we use the cross correlation to learn each node’s contribution in each hyperedge as $\mathbf{H}^m$, enhancing its representational capacity and improving the flexibility of information propagation. This is expressed as,
\begin{equation}
\mathbf{H}^m =\varepsilon\left(\Psi\left(\boldsymbol{X}^{m}\right) \Lambda\left(\boldsymbol{X}^{m}\right) \Psi\left(\boldsymbol{X}^{m}\right)^{\mathrm{T}} \Omega\left(\boldsymbol{X}^{m}\right)\right),
\end{equation}
where $\Psi\left(\boldsymbol{X}^{m}\right) \in \mathbb{R}^{T_m \times d_M}$ represents the linear transformation, $\Lambda\left(\boldsymbol{X}^{m}\right)\in \mathbb{R}^{d_m \times d_M}$ is a diagonal operation used to learn a distance metric among nodes, $\Omega\left(\boldsymbol{X}^{m}\right)\in \mathbb{R}^{T_m \times M} $ helps to determine the contribution of each node, and $\varepsilon$ is the step function.

With learned $\mathbf{H}^m$, a hypergraph convolution is performed to aggregate contextual information and thus capture long-term temporal dependencies within modalities, formulated as, 

\begin{equation}
\boldsymbol{F}_{H}^{m} = \left( \mathbf{I} - \mathbf{D}^{1 / 2} \mathbf{H}^{m} \mathbf{W}^m \mathbf{B}^{-1} \mathbf{H}^{m \mathrm{T}} \mathbf{D}^{-1 / 2} \right) \boldsymbol{X}^{m} \Theta,
\end{equation}
where $\mathbf{I}\in \mathbb{R}^{T_m \times T_m}$ is the identity matrix, $\mathbf{W}^m$ is the weight matrix, $\mathbf{D}\in \mathbb{R}^{T_m \times T_m}$ and $\mathbf{B}\in \mathbb{R}^{M \times M}$ represent the node degree matrix and the hyperedge degree matrix, respectively, $\Theta$ denotes the learnable parameters.

\vspace{-0.2cm}
\subsection{Attention-based Multimodal Graph Module (AMGM)}
In this module, we consider three modalites into a directed graph, using a graph attention mechanism to explore multimodal interaction and aggregates complementary information among different modalities. Compared with conventional undirected graph modeling approaches, directed graphs provide an explicit representation of the directionality of information flow between modalities, allowing for a more fine-grained modeling of inter-modal interaction structures and semantic influences.

\textbf{Graph Construction and Convolution.} The directed graph is defined as $\mathcal{G}_D=(\mathcal{V}_D, \mathcal{E}_D)$, where $\mathcal{V}_D=\{V,A,L\}$ denotes the node set with feature vectors $\boldsymbol{F}_{H}^{m}, m \in \{V,A,L\}$, $\mathcal{E}_D=\{(V,A), (V,L), ...,(L,V)\}$ denotes the edge set. For each edge $(i,j) \in \mathcal{E}_D$, we establish its similarity score as $A_{ij}=\operatorname{Softmax}\left(\left(\Theta_q \boldsymbol{F}_H^i\right)\left(\Theta_k \boldsymbol{F}_H^j\right)^T\right)$, which reflects the degree of attention node $i$ receives from node $j$. $\Theta_q$ and $\Theta_k$ are learnable parameters. Subsequently, for each node $i$, we aggregate all the attentive information it receives and update its representation accordingly. This is denoted as,
\begin{equation}
\begin{gathered}
\boldsymbol{F}_{D}^i=\boldsymbol{F}_{H}^i+\sum_{j \in \mathcal{N}_i} \operatorname{GAT}\left(\boldsymbol{F}_{H}^i, \boldsymbol{F}_{H}^j\right) \\
\operatorname{GAT}\left(\boldsymbol{F}_{H}^i, \boldsymbol{F}_{H}^j\right)=\sigma\left(A_{ij}-\lambda \operatorname{Mean}(A_{ij}) \mathds{1}^T\right)\left(\Theta_v \boldsymbol{F}_{H}^j\right)
\end{gathered}
\end{equation}
where $\mathcal{N}_i$ denotes the neighboring node set of node $i$, $\operatorname{GAT}$ denotes the graph attention operation, $\sigma$ is the non-linear activation $\operatorname{ReLU}$, $\mathds{1}$ is a matrix of all ones, $\operatorname{Mean}(.)$ denotes the mean operation, $\lambda$ represents a trade-off parameter that filters information with low similarities, and $\Theta_v$ are learnable parameters.

\subsection{Fusion and Prediction}
After above process, we obtain the representations $\boldsymbol{F}_{D}^m$ of three modalities with both temporal intra-modality interaction and semantic inter-modality interaction. Then, they are concatenated and passed through transformer layers for fusion, denoted as $\boldsymbol{F}=\operatorname{TransLayer}\left(\left[\boldsymbol{F}_{D}^V ; \boldsymbol{F}_{D}^A ; \boldsymbol{F}_{D}^L\right]\right)$. Finally, the fused features are fed into a classifier for emotion prediction, as $\hat{y}=\operatorname{Classifier}\left(\boldsymbol{F}\right)$, where $\hat{y}$ is the predicted sentiment.

\subsection{Feature Refinement Strategy}
To supervise the above interaction, ensuring that the model captures complementary semantics within imperfect samples, rather than being influenced by noise, we propose a feature refinement strategy that integrates knowledge from perfect samples to guide the model learning. Specifically, we pass both perfect and imperfect samples through the network to obtain representations. To differentiate, we add a superscript $^{\prime}$ to the feature representations of imperfct ones. Then, at each stage, the learned representations of imperfect samples are guided towards those of the perfect ones, to ensure positive interaction for robust learning. This is denoted as, 
\begin{equation}
\begin{gathered}
\mathcal{L}_{f}=\beta\sum_{m\in\{V,A,L\}}(\operatorname{JS}(\boldsymbol{F}_{H}^{m\prime} || \boldsymbol{F}_{H}^m)+\operatorname{JS}(\boldsymbol{F}_{D}^{m\prime} || \boldsymbol{F}_{D}^m)) \\
+(1-\beta)\operatorname{JS}(\boldsymbol{F}^{\prime} || \boldsymbol{F}),
\end{gathered}
\end{equation}
where $\operatorname{JS}$ denotes Jensen-Shannon (JS) divergence loss, used to measure the feature distribution discrepancy between the two, and $\beta$ is a trade-off parameter.  

Moreover, as a regression task, we utilize L1 loss to supervise emotion prediction, as $\mathcal{L}_{c}=L1(\hat{y},y)+L1(\hat{y}^{\prime},y)$, where $y$ is the ground-truth sentiment score. Finally, the total optimization objective of our model is expressed as $\mathcal{L}=\mathcal{L}_{c}+\lambda*\mathcal{L}_{f}$, where $\lambda$ is a trade-off parameter.

\begin{table*}[]
\centering
\caption{The AUILC value under various missing rates \{0.0, 0.1, ..., 0.9, 1.0\}. RM, TM, STM denote Random Missing, Temporal Missing and Structual Temporal Missing, respectively. The best results are highlighted with bold.}
\resizebox{\linewidth}{!}{%
\begin{tabular}{c|cccc|cccc}
\hline
\multirow{3}{*}{Method} & \multicolumn{4}{c|}{MOSI} & \multicolumn{4}{c}{MOSEI} \\ \cline{2-9} 
 & \multicolumn{1}{c|}{Clean} & RM & TM & STM & \multicolumn{1}{c|}{Clean} & RM & TM & STM \\ \cline{2-9} 
 & \multicolumn{1}{c|}{MAE } & MAE/Acc-2/F1 & MAE/Acc-2/F1 & MAE/Acc-2/F1 & \multicolumn{1}{c|}{MAE} & MAE/Acc-2/F1 & MAE/Acc-2/F1 & MAE/Acc-2/F1 \\ \hline
MulT & \multicolumn{1}{c|}{0.881} & 1.209/65.59/64.39 & 1.21/65.71/64.45 & 1.322/60.21/57.58 & \multicolumn{1}{c|}{0.559} & 0.715/68.67/68.89 & 0.715/68.52/68.7 & 0.763/61.35/61.7 \\
MISA & \multicolumn{1}{c|}{0.809} & 1.233/67.2/64.51 & 1.234/67.13/65.2 & 1.426/60.75/56.82 & \multicolumn{1}{c|}{0.571} & 0.721/73.75/73.09 & 0.72/73.77/73.09 & 0.766/71.71/69.69 \\
MAG-BERT & \multicolumn{1}{c|}{0.802} & 1.316/65.07/64.39 & 1.319/65.1/63.37 & 1.528/58.82/54.59 & \multicolumn{1}{c|}{0.536} & 0.697/74.33/74.11 & 0.698/73.48/73.55 & 0.723/70.17/69.97 \\
self-MM & \multicolumn{1}{c|}{0.79} & 1.295/67.43/65.11 & 1.295/67.65/65.41 & 1.615/60.8/56.37 & \multicolumn{1}{c|}{0.574} & 0.722/70.39/70.3 & 0.723/70.4/70.35 & 0.762/65.43/65.35 \\ \hline
T2FN & \multicolumn{1}{c|}{0.89} & 1.211/65.6/64.76 & 1.211/64.45/64.63 & 1.303/61.51/60.75 & \multicolumn{1}{c|}{0.58} & 0.723/73.27/71.63 & 0.722/73.31/71.66 & 0.76/67.72/66.24 \\
TPFN & \multicolumn{1}{c|}{0.896} & 1.195/65.23/62.67 & 1.196/65.23/62.67 & 1.267/61.41/58.58 & \multicolumn{1}{c|}{0.59} & 0.725/73.78/72.83 & 0.724/73.71/72.78 & 0.758/69.73/68.98 \\
TFR-Net & \multicolumn{1}{c|}{0.98} & 1.204/65.83/63.25 & 1.201/65.99/63.55 & 1.265/62.34/59.03 & \multicolumn{1}{c|}{0.593} & 0.725/73.39/71.44 & 0.724/73.4/71.44 & 0.756/71.28/67.74 \\
NIAT & \multicolumn{1}{c|}{\textbf{0.758}} & 1.131/68.02/66.13 & 1.13/67.95/66.06 & 1.261/61.99/58.7 & \multicolumn{1}{c|}{0.554} & 0.69/77.81/75.24 & 0.69/77.79/75.22 & 0.735/\textbf{75.29}/71.26 \\ \hline
Ours & \multicolumn{1}{c|}{0.762} & \textbf{1.097}/\textbf{69.01}/\textbf{68.05} & \textbf{1.097}/\textbf{69.01}/\textbf{68.05} & \textbf{1.193}/\textbf{63.92}/\textbf{62.34} & \multicolumn{1}{c|}{\textbf{0.528}} & \textbf{0.662}/\textbf{78.13}/\textbf{76.4} & \textbf{0.662}/\textbf{78.12}/\textbf{76.39} & \textbf{0.714}/74.81/\textbf{72.47} \\ \hline
Ours w/o TGC & \multicolumn{1}{c|}{0.75} & 1.116/68.61/67.15 & 1.116/68.61/67.14 & 1.215/62.42/59.94 & \multicolumn{1}{c|}{0.53} & 0.658/74.46/73.8 & 0.658/74.49/73.82 & 0.719/68.65/67.31 \\
Ours w/o MGC & \multicolumn{1}{c|}{0.764} & 1.116/68.69/67.39 & 1.117/68.7/67.4 & 1.214/62.96/60.14 & \multicolumn{1}{c|}{0.54} & 0.662/75.76/74.25 & 0.662/75.74/74.24 & 0.712/71.78/69.6 \\
Ours w/o strategy & \multicolumn{1}{c|}{0.802} & 1.161/67.94/66.99 & 1.161/67.95/66.98 & 1.267/61.8/59.43 & \multicolumn{1}{c|}{0.583} & 0.683/69.61/69.71 & 0.683/69.54/69.64 & 0.73/60.23/60.42 \\ \hline
\end{tabular}%
}
\end{table*}

\section{Experiments}
\subsection{Datasets and Feature Extraction}
MOSI and MOSEI are two popular benchmark datasets for MSA. The former consists of 2199 opinion video clips, with 1284, 229 and 686 clips used for train, valid and test, respectively. The latter is an improvement over MOSI made up of 22856 video clips, which has 16326, 1871 and 4659 samples in train, valid and test sets. Each sample of MOSI and MOSEI is labelled with a sentiment score ranging from -3 to +3 (strongly negative to strongly positive). For language modality, we feed the video transcripts to pre-trained Bert \cite{devlin2018bert} to obtain a 768-dimensional word embedding. For the audio modality, COVAREP toolkit \cite{degottex2014covarep} is employed to extract 74-dimensional low-level audio features. For the visual modality, the Facet is utilized to extract 35 facial action units, recording facial muscle movement and representing per-frame emotions. 

\subsection{Implementation Details and Evaluation Metrics}
The Adam optimizer is employed with a learning rate of 0.002 and 0.0001 for MOSI and MOSEI respectively. The hyper-parameter $\alpha$ and $\beta$ are adjusted from 0 to 1 with a step length of 0.1. $\lambda$ is set as 0.5. Following \cite{yuan2023noise}, the Mean Absolute Error (MAE) and Acc-2, F1 score computed for negative/non-negative classification are used as evaluation metrics. To evaluate the model's robustness against varying missing rates \{0.0, 0.1, ..., 0.9, 1.0\}, we evaluate the Area Under Indicators Line Chart (AUILC) following \cite{yuan2023noise}. Given corresponding model performance $\{e_0, e_1,...,e_n\}$ under the increasing missing rates sequence $\{r_0, r_1,...,r_n\}$, the AUILC is defined as $\sum_{i=0}^{n-1} \frac{(e_i + e_{i+1})}{2} \cdot (r_{i+1} - r_i)$. For all above metrics, higher values indicate stronger performance, except MAE where lower values indicate stronger performance. 

\subsection{Comparison With State-of-The-Art Methods}
To validate the effectiveness of our method, we compare with several state-of-the-art methods, including 1) traditional MSA methods: MuLT \cite{tsai2019multimodal}, MISA \cite{hazarika2020misa}, MAG-BERT \cite{rahman2020integrating}, self-MM \cite{yu2021learning}; 2) missing tailored MSA methods: T2FN \cite{liang2019learning}, TPFN \cite{li2020tpfn}, TFR-Net \cite{yuan2021transformer}, NIAT \cite{yuan2023noise}. The former is dedicated to employing a series of fusion techniques to integrate multimodal information, while the latter is designed for imperfect multimodal scenarios. 

\textbf{Quantitative analysis.} The quantitative results of our GIAN and baselines are shown in Table 1. We can observe that our GIAN consistently outperforms both traditional and missing tailored MSA methods across various imperfect scenarios. While traditional MSA methods perform well in learning sentiment-oriented representations for clean data, they are sensitive to missing modalities. In contrast, missing tailored methods, i.e., T2FN and TPFN devising rank regularization constraint and outer product operation respectively, mitigate sensitivity to missing features and enhance model robustness. Moreover, TFR-Net employs a transformer structure to explore multimodal correlations for reconstructing missing information, and NIAT introduces adversarial learning to recover the missing semantics. Thus, they achieve better performance compared to traditional methods under modality imperfection. However, missing tailored methods fail to sufficiently facilitate inter- and intra-modality interactions, limiting their ability to fully utilize available modalities to learn complementary semantics. Contrarily, our approach leverages graphs to mine both temporal dependencies within modalities and semantic correlations across modalities, thus achieving better robustness and generalization ability in missing scenarios. Furthermore, we notice that the model performs worse under structural temporal missing compared to random and temporal missing, which can be attributed to the increased difficulty in mining relevant missing information in such case. The improved performance of our GIAN under this case demonstrates the effectiveness of devised graph-based modules in capturing complex dependencies for complementary semantics learning. 

\begin{figure}[h]
    \centering
    \includegraphics[width=0.9\linewidth]{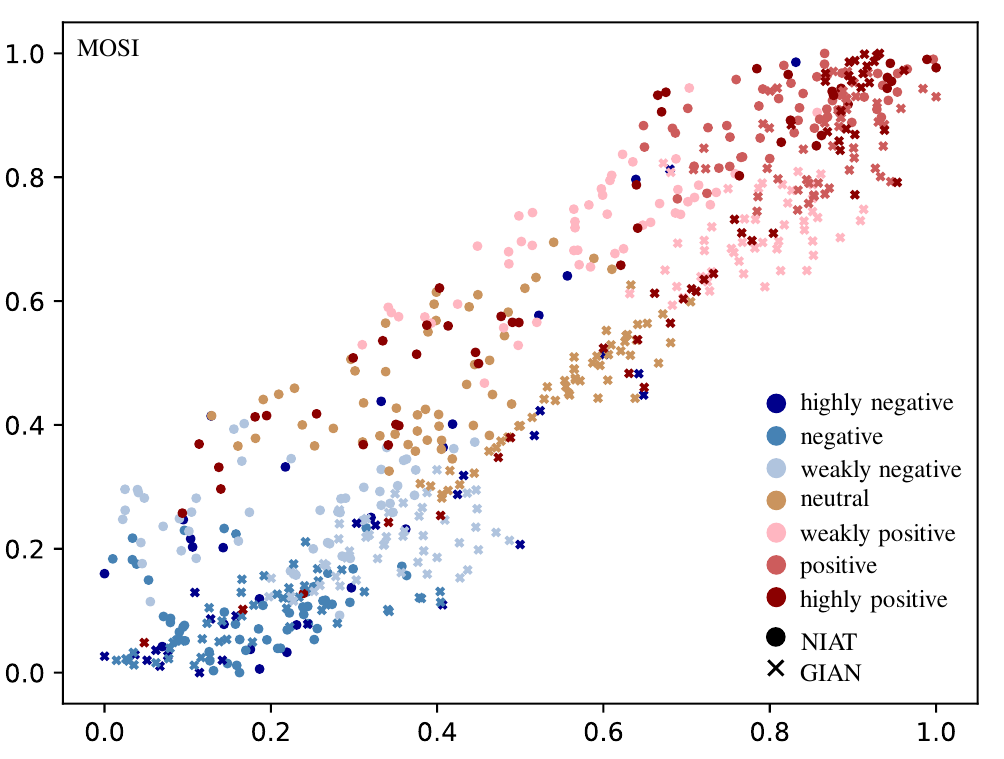}
    \caption{Visualization of joint multimodal representations on CMU-MOSI dataset, where different colors indicate sentiment intensity categories (ranging from highly negative to highly positive), and different shapes represent the feature representations produced by different models (NIAT and GIAN).} 
    \label{tsne-mosi}
\end{figure}

\begin{figure}[h]
    \centering
    \includegraphics[width=0.9\linewidth]{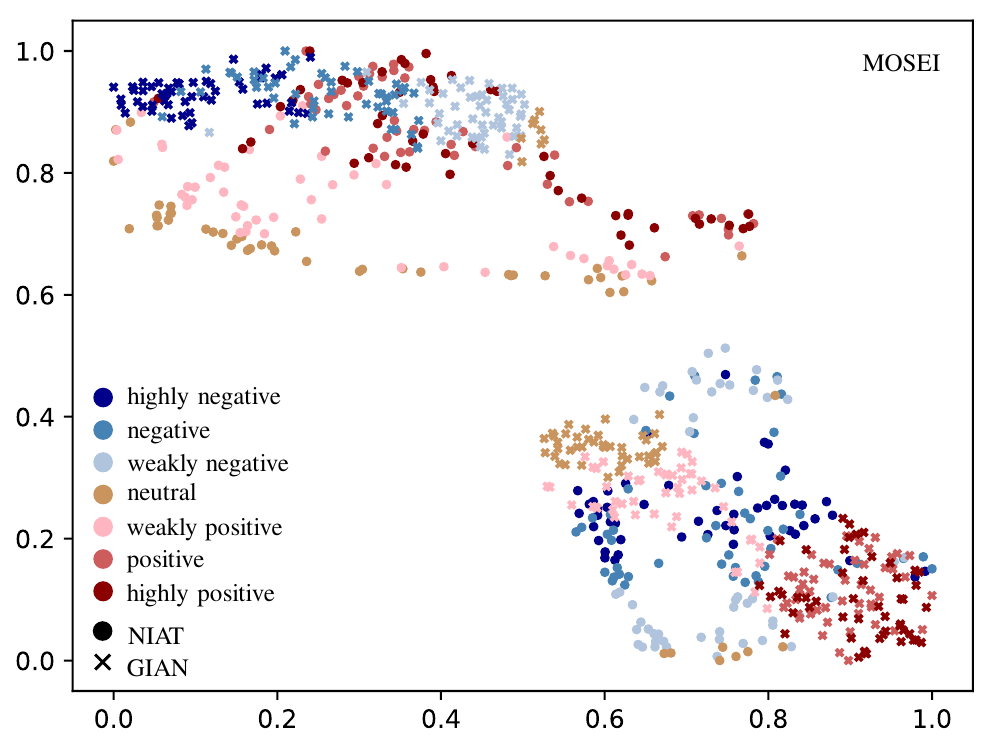}
    \caption{Visualization of joint multimodal representations on CMU-MOSEI dataset.} 
    \label{tsne-mosei}
\end{figure}

\textbf{Qualitative analysis.} To intuitively demonstrate the robustness of our method in missing scenarios, we visualize the distribution of joint representations learned by our GIAN and NIAT under 50\% temporal missing using t-SNE on CMU-MOSI and CMU-MOSEI datasets, as depicted in Fig.\ref{tsne-mosi} and Fig.\ref{tsne-mosei} respectively. The results indicate that our method learns more discriminative feature representations from incomplete data, further highlighting the effectiveness of our approach.

\subsection{Ablation Studies and Hyper-Parameter Analysis}
The last three rows of Table 1 present the ablation results of our GIAN. First, when removing the LTHM, i.e., w/o LTHM, the significantly worse performance indicates the importance of temporal intra-modality interaction for imperfect samples to capture complementary semantics within modalities. Moreover, a noteworthy drop can be observed after the removal of the AMGM (denoted as w/o AMGM), indicating that information interaction across modalities plays a crucial role in recovering missing semantics. Finally, we ablate the feature refinement strategy (denoted as w/o stratgey), i.e., only training the model using corrupted samples without explicit supervision from perfect samples, which incurs the largest performance degradation. This demonstrates the efficacy of supervision in guiding the model towards exploiting positive interactions to capture complementary semantics.

\begin{figure}[h]
    \centering
    \includegraphics[width=\linewidth]{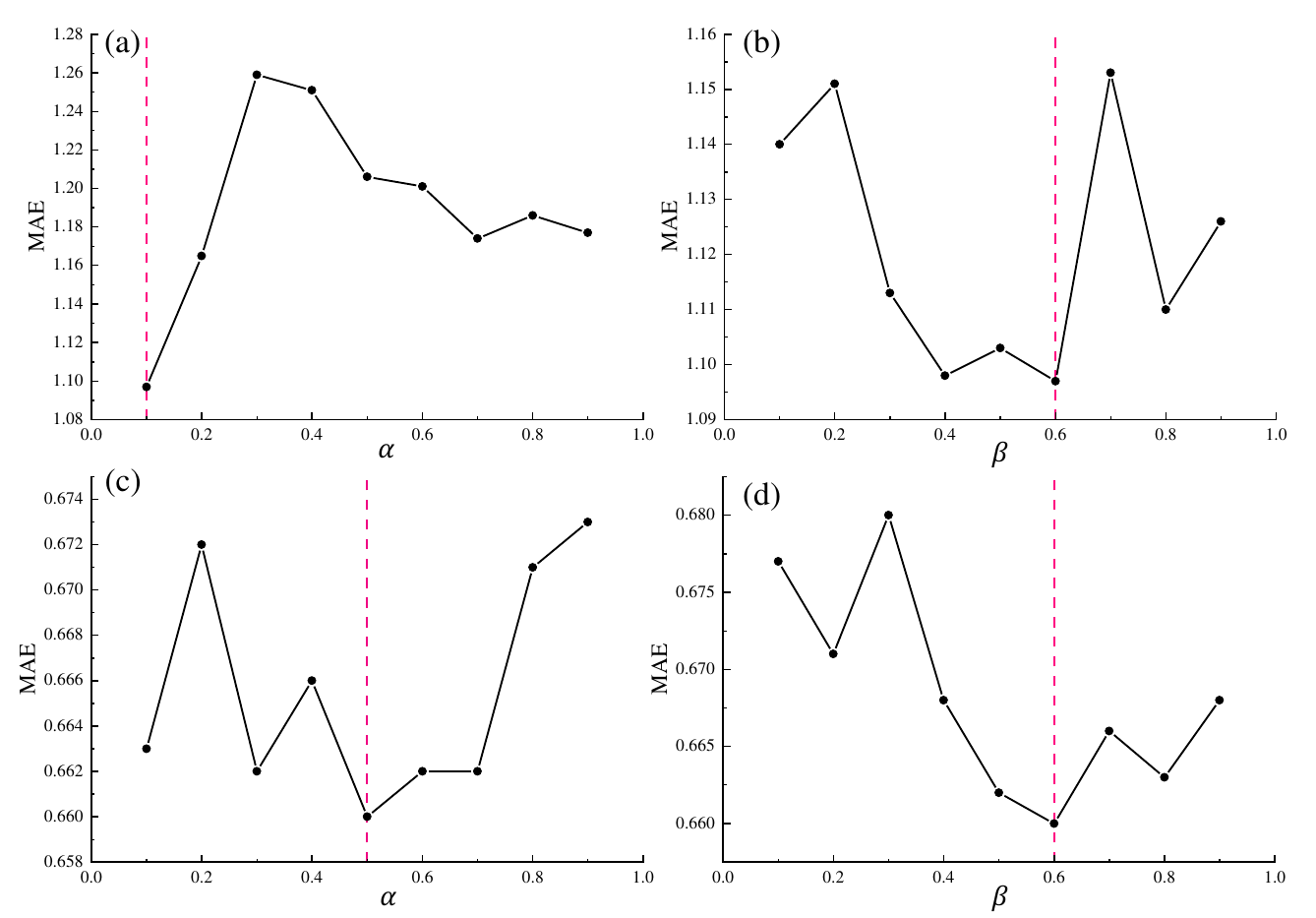}
    \caption{Performances for different $\alpha$ and $\beta$ hyper-parameters on CMU-MOSI dataset (a, b) and CMU-MOSEI dataset (c, d).} 
    \label{results}
\end{figure}

As shown in Fig.\ref{results}, we further conduct hyper-parameter
analysis on $\alpha$ and $\beta$. $\alpha$ determines the ratio of temporal missing of imperfect samples, reflecting the degree of noise introduced. The results reveal that the model gains the best performance when $\alpha$ = 0.1 and 0.5 for MOSI and MOSEI respectively. We suspect that the MOSEI dataset contains a larger number of samples with greater diversity, thus requiring more noise to train a more complex model with stronger representation capacity. Moreover, $\beta$ balances the contribution of supervision losses from different interaction stages. When $\beta$ is set to 0.6, the model achieves optimal performance. This suggests that the supervision for intra- and inter-modality interactions is relatively more important to learn robust representations. This aligns with our finding that complex interactions within and cross modalities play a crucial role for robust MSA. 

\begin{figure}[h]
    \centering
    \includegraphics[width=\linewidth]{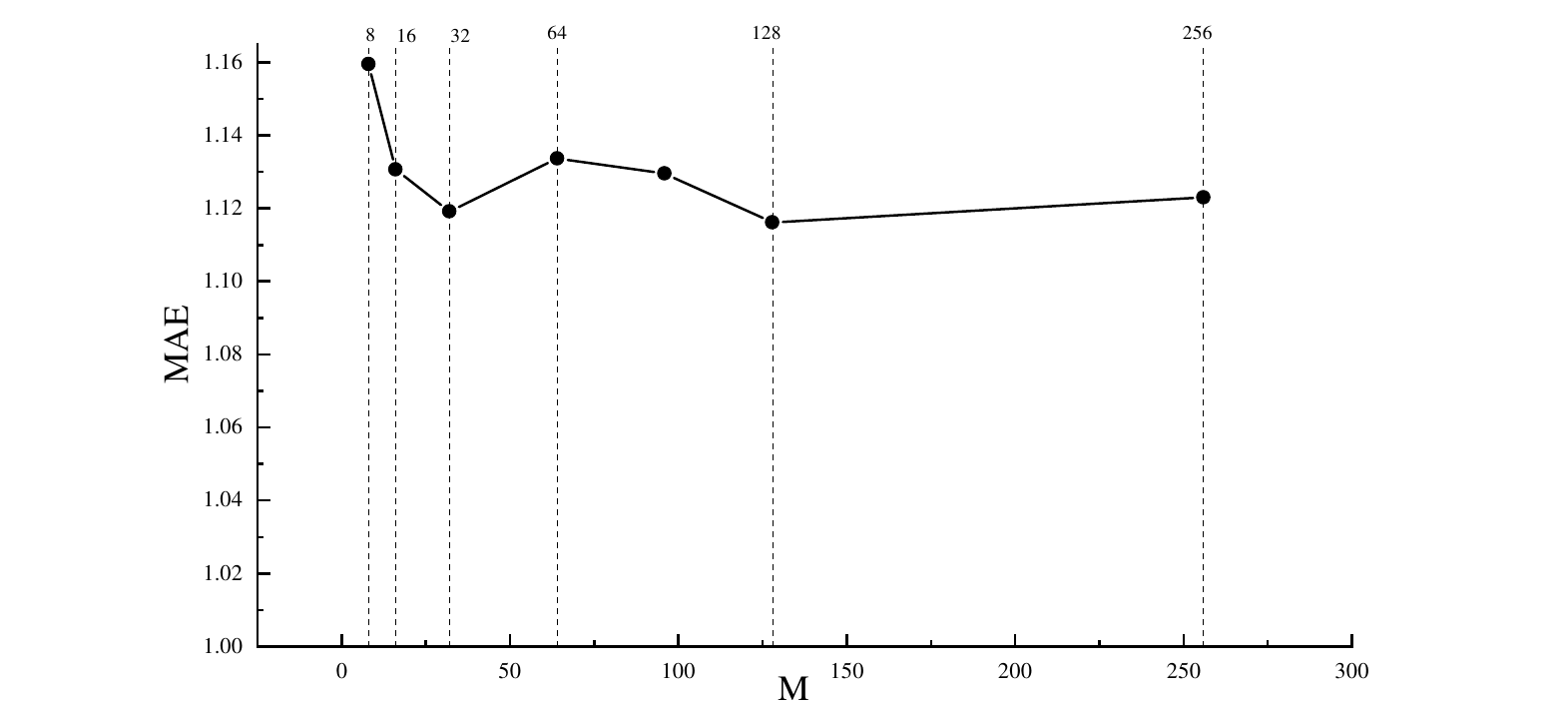}
    \caption{Performances for different $M$ on CMU-MOSI dataset.} 
    \label{results2}
\end{figure}

Finally, we investigate the impact of the number of hyperedges $M$ constructed in the LTHM module on model performance. As shown in Fig.\ref{results2}, as the number of hyperedges increases from 8 to 128, the MAE decreases initially and then tends to stabilize. This indicates that moderately increasing the number of hyperedges in the early stages can improve model performance. This is primarily because a small number of hyperedges may limit the ability to model high-order semantic relationships, while a moderate increase allows the incorporation of more potential structural information, thereby enhancing the representational capacity and robustness of the hypergraph. However, when the number of hyperedges continues to increase to 256, the MAE does not decrease significantly and even shows a slight upward trend. This suggests that an excessive number of hyperedges may introduce redundant or noisy connections, leading to a slight degradation in model performance.

\section{Conclusion}
In this paper, we propose a novel GIAN for robust MSA. Owing to devised LTHM and AMGM, our GIAN explores both temporal intra-modality dependency and semantic intermodality interaction to enable noisy samples to capture complementary semantics. A feature refinement strategy is proposed, with knowledge from perfect samples integrated to supervise the interaction process, further ensuring the model learns reliable and robust representations. Extensive experiments reveal that our GIAN yields better generalization and robustness under various modality imperfections.

\bibliographystyle{IEEEtran}
\bibliography{IEEEabrv,bare_jrnl}
\end{document}